\documentclass[superscriptaddress,twocolumn,10pt]{revtex4-1}

\usepackage{amsmath}    
\usepackage{graphicx}   
\usepackage{verbatim}   
\usepackage{color}      
\usepackage{subfigure}  
\usepackage{hyperref}   
\usepackage{amssymb}
\usepackage{amsthm}
\usepackage{amsfonts}
\begin{document}


\title{Relation Between Pore Size and the Compressibility of a Confined Fluid}

\author{Gennady Y. Gor} \email[Corresponding author, e-mail: ]{gennady.y.gor@gmail.com} 
\affiliation{NRC Research Associate, Resident at Center for Computational Materials Science, Naval Research Laboratory, Washington, DC 20375, USA}
\author{Daniel W. Siderius} 
\affiliation{Chemical Sciences Division, National Institute of Standards and Technology, Gaithersburg, MD 20899, USA}
\author{Christopher J. Rasmussen}
\affiliation{DuPont Central Research and Development Experimental Station, Wilmington, DE 19803, USA}
\author{William P. Krekelberg}
\affiliation{Chemical Sciences Division, National Institute of Standards and Technology, Gaithersburg, MD 20899, USA}
\author{Vincent K. Shen}
\affiliation{Chemical Sciences Division, National Institute of Standards and Technology, Gaithersburg, MD 20899, USA}
\author{Noam Bernstein}
\affiliation{Center for Computational Materials Science, Naval Research Laboratory, Washington, DC 20375, USA}

\date{\today}

\begin{abstract}
\noindent When a fluid is confined to a nanopore, its thermodynamic properties differ from the properties of a bulk fluid, so measuring such properties of the confined fluid can provide information about the pore sizes. Here we report a simple relation between the pore size and isothermal compressibility of argon confined in these pores. Compressibility is calculated from the fluctuations of the number of particles in the grand canonical ensemble using two different simulation techniques: conventional grand-canonical Monte Carlo and grand-canonical ensemble transition-matrix Monte Carlo. Our results provide a theoretical framework for extracting the information on the pore sizes of fluid-saturated samples by measuring the compressibility from ultrasonic experiments. 

\end{abstract}

\maketitle

\section{Introduction}

Confinement of fluid in a small pore significantly alters the fluid's thermodynamic properties. Here we focus on the fluid compressibility, which is one of the key parameters governing the mechanical properties of fluid-saturated porous materials. For example, the knowledge of the compressibility of a fluid in the nanopores is necessary for predicting the flow of hydrocarbons in tight geological formations such as shale gas. Recent studies in characterization of shales show that they typically have a substantial amount of pores that are only a few nanometers wide (mesopores) \cite{Loucks2009, Chalmers2012, Kuila2013}.  Therefore, a fundamental understanding of the effect of nano-confinement on fluid properties, such as the compressibility, is of major importance in the development of modern petroleum extraction techniques. In addition, as we show here, the dependence of fluid compressibility on pore size may be the basis for a new method for experimentally determining pore sizes.

The compressibility of a fluid confined in the nanopores can be determined experimentally from measurements of the speed of sound propagation through a saturated porous sample. Such experiments for argon and hexane in the pores of Vycor glass show that the fluid compressibility is affected by the Laplace pressure in the pores~\cite{Page1993, Page1995, Schappert2008, Schappert2014}. This effect has been recently confirmed by calculations based on macroscopic thermodynamics and classical  density functional theory (DFT)~\cite{Gor2014}.  Additionally, the calculations from Ref.~\onlinecite{Gor2014} predict that compressibility of the fluid confined in the pore linearly decreases with decreasing pore size. However, these results are based on macroscopic theories and, therefore, applying them to the pores below ca.\ 10~nm in size is questionable. 

Here we calculate the compressibility of a confined fluid as a function of pore size without exploiting any macroscopic assumptions. We use conventional grand-canonical Monte Carlo (GCMC) simulations~\cite{Norman1969} to calculate the compressibility of Lennard-Jones (LJ) liquid argon in silica pores from the fluctuations of number of particles. Since conventional GCMC is known to be inefficient for dense fluids~\cite{Frenkel-Smit}, we also perform simulations using the grand-canonical transition-matrix Monte Carlo method (GC-TMMC)~\cite{Siderius2013}. Argon was chosen for two reasons: it is simple to model, since a LJ pair potential model accurately captures its relevant interactions and geometry, and it is widely used in physisorption experiments for characterization of porous materials~\cite{Thommes2014}.  Both methods confirm the linear relation between the compressibility and the pore size from Ref.~\onlinecite{Gor2014} in the range of the pore sizes from 2.5~nm to 6.0~nm.

Our results provide a simple linear relation between the confined fluid compressibility and the pore size. The compressibility can be calculated from the speed of sound measured in ultrasonic experiments, and therefore, our study provides a theoretical framework for determination of pore sizes from ultrasonic experiments on fluid saturated samples. Such measurement can be used as a way to get the pore size distribution of a sample, alternatively to the conventional method based on the adsorption isotherm~\cite{Thommes2014}.

\section{Method}

We focus here on calculations of the isothermal compressibility $\beta_T$, defined as
\begin{equation}
\label{beta-def}
\beta_T \equiv - \frac{1}{V} \left( \frac{\partial V}{\partial P}\right)_{T},
\end{equation}
where $V$ is the volume of the pore, $P$ is the fluid pressure, and $T$ is the absolute temperature. 
When the fluid is confined in a nanopore, the interaction with confining walls may introduce anisotropy and inhomogeneity in the system, so that instead of a single scalar pressure value one has to introduce a pressure tensor, and it may vary as a function of position relative to the pore wall.  This pressure tensor for the simple fluids in the nanopores can be calculated using asymptotic theories \cite{Brodskaya2010i,Brodskaya2010ii} or molecular simulations \cite{Long2011,Long2013,Long2013Colloids}. One could also introduce a spatially-dependent compressibility tensor, defined similarly to Eq.~\ref{beta-def}, but using components of the pressure tensor. However, it is unlikely that the anisotropy and heterogeneity of the fluid compressibility in the pore can be measured by ultrasonic experiments. Instead, we calculate the scalar compressibility for the fluid in the pore as a whole, which is likely to represent the experimentally observable, average macroscopic value. Note that our calculation of the compressibility does not require the calculation of pressure.

Classical statistical mechanics allows the calculation of the compressibility of the fluid $\beta_T$ from the fluctuations of number of particles in the pore $N$ in the grand-canonical ensemble~\cite{Landau5, Allen1989} through the relation
\begin{equation}
\label{beta-fluct}
\beta_T = \frac{V \langle \delta N^2 \rangle}{k_{\rm{B}} T \langle N\rangle^2 },
\end{equation}
where $\langle \delta N^2 \rangle = \langle N^2 \rangle - \langle N \rangle^2$ and $k_{\rm{B}}$ is Boltzmann's constant. Note that Eq.~\ref{beta-fluct} is valid when the fluctuations are normally distributed~\cite{Landau5}. In the grand-canonical ensemble the fluid atoms or molecules in the pore are assumed to be in equilibrium with a reservoir, and the pressure $p$ in the reservoir is directly related to the chemical potential $\mu$. In a typical adsorption experiment the reservoir contains gas phase and the pressure $p$ in the gas phase, is different from the pressure in the adsorbed phase $P$.

\subsection{Conventional Grand Canonical Monte Carlo Simulations}

We modeled argon at its normal boiling point $T = 87.3$~K in spherical silica pores of different sizes. We began by performing conventional GCMC simulations~\cite{Norman1969}, based on the Metropolis algorithm~\cite{Metropolis1953}. This approach is widely used for modeling fluids adsorption in nanopores~\cite{Gubbins2011}. Interactions between the argon atoms are modeled by the LJ pair potential. To model the attractive adsorption potential between the fluid and pore walls we used a spherically integrated, site-averaged LJ potential~\cite{Baksh1991, Ravikovitch2002}. The parameters for intermolecular potentials are summarized in Table~\ref{sim-parameters}.

\begin{table}[ht]
  \begin{center}
  \begin{tabular}{| c | c | c | c | c | c | }
  \hline
  Interaction & $\sigma$, nm & $\epsilon/k_B$, K & $\rho_s$, nm$^{-2}$ & $r_{\rm{cut}}$, $\sigma_{\rm{ff}}$ & Ref \\
  \hline
  Ar-Ar & 0.34 & 119.6 & - & 5 & \cite{Vishnyakov2001} \\
  Silica-Ar & 0.30 & 171.24 & 15.3 & 10 & \cite{Ravikovitch1997} \\  
    \hline
  \end{tabular}
  \caption{Lennard-Jones parameters and relevant physical properties for the Ar-Ar fluid-fluid (ff) interaction and Silica-Ar solid-fluid (sf) interaction. $\sigma$ is the LJ diameter, $\epsilon$ is the LJ energy scale, $\rho_s$ is the surface number density of solid LJ sites, and $r_{\rm{cut}}$ is the distance at which the interactions were truncated; no tail corrections are used.
    \label{sim-parameters}
}
  \end{center}
\end{table}

The calculations were performed for a number of different pore sizes from 2.5~nm to 6.0~nm. By the pore sizes we refer here to the external diameter of a spherical pore $d_{\mathrm{ext}}$, the distance of a line drawn through the centers of hypothetical silica solid atoms at opposite pore wall surfaces. When the volume $V$ of the pore accessible by the fluid atoms is calculated (used e.g.,\ in Eqs.~\ref{beta-def}, \ref{beta-fluct}), a different diameter is used: internal diameter determined from the root of the spherically-integrated solid-fluid potential. For pores $d_{\mathrm{ext}} \gtrsim 1.5$~nm, the internal diameter is given by $d_{\mathrm{int}} \simeq d_{\mathrm{ext}} - 1.7168 \sigma_{\rm{sf}} + \sigma_{\rm{ff}}$~\cite{Rasmussen2010, Gor2012}. 

All the simulations were performed at the chemical potential $\mu^* = -9.6 \epsilon_{\rm{ff}}$ , which correspond to the saturation pressure $p_0$ of LJ fluid at the considered temperature, calculated from the Johnson et al. equation of state~\cite{Johnson1993}. Hereinafter the quantities marked with an  asterisk are presented in reduced (LJ) units.

When GCMC is performed in liquid-like states, the number of successful insertions is low, thus in order to achieve a well equilibrated initial state, at each pore size a GCMC simulation was run for at least  $10^{9}$ trial MC moves. Then three different configurations from this trajectory were picked and  used as initial conditions for three independent MC trajectories with different random seeds, $5 \times 10^{9}$ trial moves each. The results of these three trajectories were used for calculations of compressibility.  The compressibility was calculated using Eq.~\ref{beta-fluct}.  To estimate the errors for the calculated compressibility, the method based on the autocorrelation function was used \cite{Note1}.

\subsection{Transition-Matrix Monte-Carlo}

Since the conventional GCMC method is known to be problematic for simulating dense fluids, we also performed molecular simulations using the GC-TMMC method. The GC-TMMC method has been described in detail by Errington and coworkers~\cite{Errington_Evaluating_2003,Errington_Direct_2003,Shen_Metastability_2004,Shen_Determination_2005,Errington_Direct_2005,Shen_Determination_2006} and has been specifically discussed in the context of gas adsorption in porous materials~\cite{Siderius2013,Shen_Elucidating_2014}. Here we use an implementation of GC-TMMC identical to that in 
Ref.~\onlinecite{Siderius2013}, except that the simulations were initialized using the Wang-Landau algorithm to pre-fill the TMMC collection matrix~\cite{Shell_improved_2003,Rane_Monte_2013,Shen_Elucidating_2014}.

The main advantage of GC-TMMC over conventional GCMC is that it has much lower statistical noise by using the full transition matrix to estimate ensemble averages.  In addition, the implementation we use takes advantage of the methods of expanded ensembles~\cite{Lyubartsev_New_1991,Escobedo_Expanded_1996} and windowed $N$ domain~\cite{Errington_Prewetting_2004} to further improve performance.  As implemented, GC-TMMC utilized a bias function to efficiently sample the entire relevant range of particle numbers $N$ at each $\mu$, thus making it possible to use histogram reweighting \cite{Errington_Evaluating_2003, Errington_Direct_2003} to accurately calculate results at values of $\mu$ that are different from the one that was explicitly simulated. Using this approach it is possible to efficiently calculate the entire adsorption isotherm, i.e.  the adsorbed fluid density $n=N/V$ as a function of $\mu$.

In the present work, two separate GC-TMMC simulations were run. First, we simulated bulk, unconfined, argon at $T = 87.3$~K to 
obtain the equation of state (i.e., the relationship between $\mu$ and $p$), the saturation pressure, the and compressibility $\beta_ T$ of the bulk liquid. Second, we simulated argon in spherical silica pores at various pore sizes (2.5~nm, 3.0~nm, 4.0~nm, and 5.0~nm), with the same ff and sf potential as described above (Table~\ref{sim-parameters}). Each GC-TMMC simulation was run for $2 \times 10^{10}$ trial moves, of which 40\% were displacements and the remainder were exchange moves, and used to calculated $\left\langle N\right\rangle$ and $\beta_T$ at a range of desired values of $\mu$ (or equivalently reservoir pressures $p$).  Lastly, the adsorption isotherm $n(\mu)$ was obtained using the procedure in Ref.~\onlinecite{Siderius2013}.  Uncertainty in calculated properties was estimated using t-statistics based on four independent simulations.

\subsection{Macroscopic Approach}

As an alternative to the statistical mechanics expression of Eq.~\ref{beta-fluct}, compressibility can be calculated from the slope of the adsorption isotherm $n(p/p_0)$, following our previous work~\cite{Gor2014}. Assuming that the fluid confined in the pore is a uniform macroscopic system, we can relate the pressure in the pore $P$ to the chemical potential of the fluid $\mu$ through the Gibbs-Duhem relation at constant temperature
\begin{equation}
\label{Gibbs-Duhem}
d P = n ~ d \mu.
\end{equation} 
Using $n=N/V$ and Eq.~\ref{Gibbs-Duhem}, Eq.~\ref{beta-def} can be rewritten as 
\begin{equation}
\label{beta-bulk}
\beta_T = \frac{1}{n^2} \left( \frac{\partial n}{\partial \mu}\right)_{T,N} = \frac{1}{n^2} \left( \frac{\partial n}{\partial \mu}\right)_{T,V}.
\end{equation}
Since $n$ is an intensive variable, it depends on $\mu$ and $T$ and does not depend on the extensive variables $N$ and $V$. It allows us to change the derivative at $V=\rm{const}$ to $N=\rm{const}$ in Eq.~\ref{beta-bulk}~\cite{Kuni1981} . The last expression in the right hand side of Eq.~\ref{beta-bulk} can be calculated by numerical differentiation of the adsorption isotherm.

\section{Results}

\begin{figure}
\includegraphics[width=0.95\linewidth]{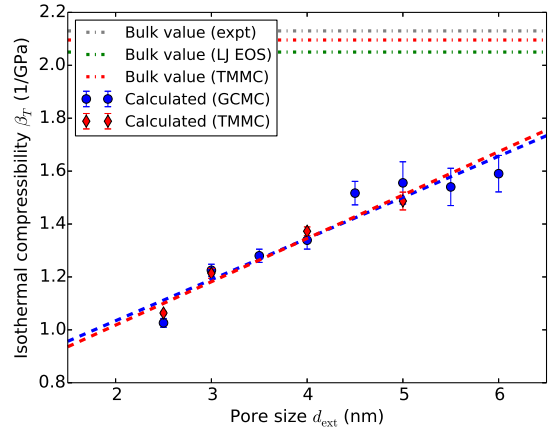}
\caption{Isothermal compressibility $\beta_T$ of liquid argon at $T=87.3$~K at saturation ($p/p_0 = 1$) as a function of pore diameter $d_{ext}$. Horizontal dash-dotted lines give the values for bulk argon based on experimental data from~\cite{Tegeler1999}, on the equation of state of the LJ fluid~\cite{Johnson1993}, and obtained from GC-TMMC simulations for bulk argon.  Compressibility calculated using Eq.~\ref{beta-fluct} is shown by symbols for GCMC (circles) and GC-TMMC (diamonds).  Error bars for GCMC are estimated based on \cite{Note1}; 
for GC-TMMC error bars based on t-statistics and independent simulations are smaller than the symbol size for diameters less than 5~nm.  Dashed lines show linear fits for each method.}
\label{fig-gcmc}
\end{figure}

For bulk LJ argon, compressibility can be calculated using Eq.~\ref{beta-bulk} from the Johnson EOS~\cite{Johnson1993}, which for  $T = 87.3$~K at the saturation point gives $\beta_T = 2.05$ GPa$^{-1}$. Note that the Johnson EOS is based on GCMC simulations. This number is close to the experimental value for Ar: the bulk data from Table 33 of Ref.~\cite{Tegeler1999}, interpolated to $T = 87.3$~K, give $\beta_T = 2.13$ GPa$^{-1}$. GC-TMMC simulations give the following properties for the bulk argon: saturation pressure $p_0 = 0.9298 \pm 0.0002~\rm{bar}$, chemical potential $\mu_{0}^* = -9.6024 \pm 0.0002$, and isothermal compressibility  $\beta_T = 2.104 \pm 0.046~\rm{GPa}^{-1}$. These three different bulk compressibility values are shown in Fig.~\ref{fig-gcmc}.

Figure~\ref{fig-gcmc} presents results of calculation of the isothermal compressibility using both GCMC and GC-TMMC methods; both methods show a clear linear trend for $\beta_T$ as a function of the pore size, confirming the result of Ref.~\onlinecite{Gor2014}. Even for relatively large mesopores (6 nm in diameter), the calculated compressibility is noticeably lower than the bulk compressibility. For the smallest pore size considered here (2.5~nm in diameter), the compressibility is 1.06~GPa$^{-1}$, which is about 2 times lower than the bulk value.  

\begin{figure}
\includegraphics[width=0.95\linewidth]{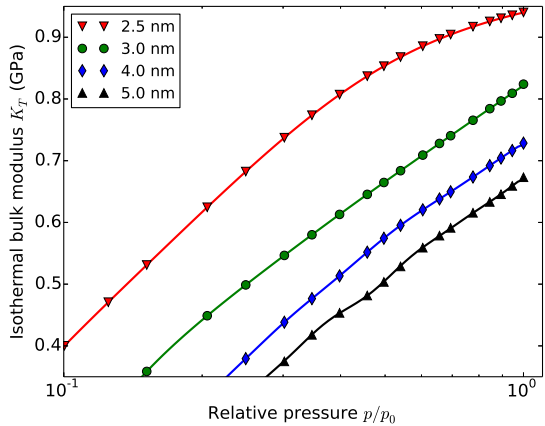}
\caption{Isothermal bulk modulus ($K_T = 1/\beta_T$) of liquid argon at $T=87.3$~K confined in the pores of various sizes as a function of gas pressure $p$ (log scale). The solid lines give the data based on the macroscopic approach, derived from GC-TMMC isotherms (Eq.~\ref{beta-bulk}), the symbols present the data obtained directly from the statistical mechanics expression Eq.~\ref{beta-fluct}.}
\label{fig-pressure}
\end{figure}

Performing the histogram reweighting on the results of GC-TMMC for a range of chemical potentials, we get a high resolution adsorption isotherm $n = n(\mu) = n(p/p_0)$, which can be numerically differentiated to calculate the compressibility for a given pore as a function of $p/p_0$ (using Eq.~\ref{beta-bulk}). Figure~\ref{fig-pressure} shows the isothermal bulk modulus of the confined fluid $K_T \equiv 1/\beta_T$ calculated using Eq.~\ref{beta-fluct} (points) and using Eq.~\ref{beta-bulk} based on GC-TMMC simulations (lines) for 2.5~nm, 3~nm, 4~nm and 5~nm pores~\cite{Note2}. Our calculations show that the dependence of $K_T$ on $\log(p/p_0)$ is nearly linear, which is in agreement with the previous DFT calculations~\cite{Gor2014} and experimental observations~\cite{Page1995, Schappert2014}. 

\section{Discussion}

The fact that the compressibility of a bulk fluid differs from the compressibility of the same fluid in confinement has been reported earlier by Coasne and coauthors~\cite{Coasne2009} based on GCMC simulations for LJ argon confined in a slit micropore. However, to our knowledge, the pore-size dependence of the compressibility has not been investigated systematically. The physical reason for this dependence is straightforward: the attractive walls lead to formation of fluid layers with high density near them. These dense layers have been observed recently for water adsorption in silica pores by combination of \textit{in situ} X-ray and neutron scattering \cite{Erko2012}. For small pores these layers contribute significantly to the average density of the confined fluid, so that it becomes noticeably higher than the bulk density at the same chemical potential. Since the liquid compressibility has a strong inverse dependence on density,  the confined fluid has lower compressibility than bulk. 

Also, the calculation of pressure in confined fluids shows that even at saturation ($p = p_0$), the pressure $P$ in the fluid is $P \gg p_0$~\cite{Gor2010}, so that a confined fluid is effectively under compression due to the attraction of the pore walls. Obviously, the confined fluid, which is already ``compressed'' at a high pressure $P$, has lower compressibility then the one which has not been compressed, i.e. the bulk.

Our molecular simulations go beyond this qualitative conclusion, and can provide quantitative values of compressibility for a fluid confined in pores of different sizes. The methods we use here were verified previously for predicting thermodynamic properties of confined fluids, and as we have shown here, in the limit of large pores they give fluid compressibility values that are in good agreement with published bulk data.

In this work we present compressibility predictions for LJ argon in mesopores with diameters ranging from 2.5~nm to 5.0~nm (GC-TMMC) or 6.0~nm (GCMC). For pores larger than that, both methods become too computationally expensive. However, extrapolating the linear trend in Fig.~\ref{fig-gcmc} leads to an intersection with the bulk horizontal line at 9.0~nm (GCMC) or 8.8~nm (GC-TMMC). It is likely that for pore sizes larger than that, the compressibility will be close to the bulk value. For pores smaller than 2.5~nm both GCMC and GC-TMMC methods give particle number distributions that are essentially non-Gaussian. Therefore, Eq.~\ref{beta-fluct} for calculation of compressibility is not applicable. Moreover, the use of a macroscopic concept of compressibility for such small systems ($\langle N \rangle \lesssim 100$) is questionable.

Note that as Fig.~\ref{fig-pressure} shows, for all the pore sizes considered calculations based on the macroscopic expression in Eq.~\ref{beta-bulk} are in excellent agreement with the result from Eq.~\ref{beta-fluct}, which do not use any macroscopic assumptions. This agreement shows that the Gibbs-Duhem equation (Eq.~\ref{Gibbs-Duhem}) remains valid even for a pore as small as of 2.5~nm diameter, which has $\langle N \rangle = 125$.

We do not present a direct comparison to experimental data, because to our knowledge there has been no systematic study of the compressibility of confined argon in pores of different sizes. Such a study seems feasible, since the method for calculation of the compressibility of confined fluids from ultrasonic experiments is well-developed~\cite{Warner1988, Page1995, Schappert2014}. Note that for direct comparison between compressibility data obtained from simulation and ultrasonic experiment, one has to take into account that the former is an isothermal process while the latter is adiabatic, described in terms of adiabatic compressibility $\beta_S =  {\beta_T}/{\gamma}$, where $\gamma$ is the heat capacity ratio of the fluid $\gamma \equiv c_P/c_V$. The parameter $\gamma$ is not sensitive to pressure~\cite{Tegeler1999}, and consequently, $\gamma$ for a confined fluid should not be much different from the bulk. Therefore, following~\cite{Gor2014} for comparison of the presented data to the experimental data from ultrasonic experiments one can use $\gamma = 2.1$ for bulk liquid argon from Ref.~\onlinecite{Tegeler1999}.

A relation between an experimentally-measurable thermodynamic property of a fluid in the pore and the pore size can be used for determination the latter. A conventional adsorption experiment in a mesopore is a typical example: measuring the values of gas pressures at which the capillary condensation and evaporation in a porous material take place allows one to estimate the pore size and even to calculate the pore-size distribution (PSD). This is the basis of porosimetry via the classical Barrett-Joyner-Halenda (BJH) method~\cite{Barrett1951} and more advanced methods based on classical DFT~\cite{Landers2013}.  We propose that the unambiguous relation between the fluid compressibility and the pore size we have calculated can also be used as a basis for a method to estimate of the pore sizes of fluid-filled nanoporous samples from the ultrasonic experiments. Such a method would be different from the ultrasonic methods proposed before by Dukhin et al.~\cite{Dukhin2013} and by Warner and Beamish~\cite{Warner1988}. The former is based on the measurement of seismoelectric current and provides the information on the PSD in the macropore ($d > 50$~nm) regime; the latter measures the {\em mass} of the adsorbed fluid. Therefore, our study provides the first critical steps toward an application of technological relevance, offering a basis for a totally new approach for determination of pore sizes of nanoporous materials. One particular advantage of our proposed method for ultrasonic porosimetry is that it can be done \textit{in situ}, allowing pore size determination in a wide range of working fluids and experimental conditions.

The physical reason for the relation between compressibility and pore size is not specific to LJ interactions. We expect that a similar relation can be observed for more complex systems, e.g.,\ confined water, carbon dioxide or hydrocarbons, which are attracting much practical interest~\cite{Mosher2013, Hu2015, Falk2015}. Consequently, our study sheds light beyond the proposed technological application, enhancing our fundamental understanding of the properties of nano-confined fluids.

\section{Conclusion}

To summarize, when a fluid is confined to a nanopore, its thermodynamic properties differ from the properties of a bulk fluid. Measurement of such properties of a confined fluid can provide useful information about the pore sizes. Here we report a simple relation between the pore size and isothermal compressibility of argon confined in these pores. Compressibility is calculated from the fluctuations of the number of particles in the grand canonical ensemble using two different simulation techniques, conventional GCMC and GC-TMMC. Our results provide a theoretical framework for extracting the information on the pore sizes of fluid-saturated samples from measurements of the compressibility from ultrasonic experiments. 

\section*{Acknowledgments}

This research was performed while one of the authors (G.G.) held a National Research Council Research Associateship Award at Naval Research Laboratory. The work of G.G. and N.B. was funded by the Office of Naval Research through the Naval Research Laboratory's basic research program. 


\begin{thebibliography}{56}
\expandafter\ifx\csname natexlab\endcsname\relax\def\natexlab#1{#1}\fi
\expandafter\ifx\csname bibnamefont\endcsname\relax
  \def\bibnamefont#1{#1}\fi
\expandafter\ifx\csname bibfnamefont\endcsname\relax
  \def\bibfnamefont#1{#1}\fi
\expandafter\ifx\csname citenamefont\endcsname\relax
  \def\citenamefont#1{#1}\fi
\expandafter\ifx\csname url\endcsname\relax
  \def\url#1{\texttt{#1}}\fi
\expandafter\ifx\csname urlprefix\endcsname\relax\def\urlprefix{URL }\fi
\providecommand{\bibinfo}[2]{#2}
\providecommand{\eprint}[2][]{\url{#2}}

\bibitem[{\citenamefont{Loucks et~al.}(2009)\citenamefont{Loucks, Reed, Ruppel,
  and Jarvie}}]{Loucks2009}
\bibinfo{author}{\bibfnamefont{R.~G.} \bibnamefont{Loucks}},
  \bibinfo{author}{\bibfnamefont{R.~M.} \bibnamefont{Reed}},
  \bibinfo{author}{\bibfnamefont{S.~C.} \bibnamefont{Ruppel}},
  \bibnamefont{and} \bibinfo{author}{\bibfnamefont{D.~M.}
  \bibnamefont{Jarvie}}, \bibinfo{journal}{J. Sediment. Res.}
  \textbf{\bibinfo{volume}{79}}, \bibinfo{pages}{848} (\bibinfo{year}{2009}).

\bibitem[{\citenamefont{Chalmers et~al.}(2012)\citenamefont{Chalmers, Bustin,
  and Power}}]{Chalmers2012}
\bibinfo{author}{\bibfnamefont{G.~R.} \bibnamefont{Chalmers}},
  \bibinfo{author}{\bibfnamefont{R.~M.} \bibnamefont{Bustin}},
  \bibnamefont{and} \bibinfo{author}{\bibfnamefont{I.~M.} \bibnamefont{Power}},
  \bibinfo{journal}{AAPG Bulletin} \textbf{\bibinfo{volume}{96}},
  \bibinfo{pages}{1099} (\bibinfo{year}{2012}).

\bibitem[{\citenamefont{Kuila and Prasad}(2013)}]{Kuila2013}
\bibinfo{author}{\bibfnamefont{U.}~\bibnamefont{Kuila}} \bibnamefont{and}
  \bibinfo{author}{\bibfnamefont{M.}~\bibnamefont{Prasad}},
  \bibinfo{journal}{Geophys. Prospect.} \textbf{\bibinfo{volume}{61}},
  \bibinfo{pages}{341} (\bibinfo{year}{2013}).

\bibitem[{\citenamefont{Page et~al.}(1993)\citenamefont{Page, Liu, Abeles,
  Deckman, and Weitz}}]{Page1993}
\bibinfo{author}{\bibfnamefont{J.~H.} \bibnamefont{Page}},
  \bibinfo{author}{\bibfnamefont{J.}~\bibnamefont{Liu}},
  \bibinfo{author}{\bibfnamefont{B.}~\bibnamefont{Abeles}},
  \bibinfo{author}{\bibfnamefont{H.~W.} \bibnamefont{Deckman}},
  \bibnamefont{and} \bibinfo{author}{\bibfnamefont{D.~A.} \bibnamefont{Weitz}},
  \bibinfo{journal}{Phys. Rev. Lett.} \textbf{\bibinfo{volume}{71}},
  \bibinfo{pages}{1216} (\bibinfo{year}{1993}).

\bibitem[{\citenamefont{Page et~al.}(1995)\citenamefont{Page, Liu, Abeles,
  Herbolzheimer, Deckman, and Weitz}}]{Page1995}
\bibinfo{author}{\bibfnamefont{J.~H.} \bibnamefont{Page}},
  \bibinfo{author}{\bibfnamefont{J.}~\bibnamefont{Liu}},
  \bibinfo{author}{\bibfnamefont{B.}~\bibnamefont{Abeles}},
  \bibinfo{author}{\bibfnamefont{E.}~\bibnamefont{Herbolzheimer}},
  \bibinfo{author}{\bibfnamefont{H.~W.} \bibnamefont{Deckman}},
  \bibnamefont{and} \bibinfo{author}{\bibfnamefont{D.~A.} \bibnamefont{Weitz}},
  \bibinfo{journal}{Phys. Rev. E} \textbf{\bibinfo{volume}{52}},
  \bibinfo{pages}{2763} (\bibinfo{year}{1995}).

\bibitem[{\citenamefont{Schappert and Pelster}(2008)}]{Schappert2008}
\bibinfo{author}{\bibfnamefont{K.}~\bibnamefont{Schappert}} \bibnamefont{and}
  \bibinfo{author}{\bibfnamefont{R.}~\bibnamefont{Pelster}},
  \bibinfo{journal}{Phys. Rev. B} \textbf{\bibinfo{volume}{78}},
  \bibinfo{pages}{174108} (\bibinfo{year}{2008}).

\bibitem[{\citenamefont{Schappert and Pelster}(2014)}]{Schappert2014}
\bibinfo{author}{\bibfnamefont{K.}~\bibnamefont{Schappert}} \bibnamefont{and}
  \bibinfo{author}{\bibfnamefont{R.}~\bibnamefont{Pelster}},
  \bibinfo{journal}{EPL} \textbf{\bibinfo{volume}{105}}, \bibinfo{pages}{56001}
  (\bibinfo{year}{2014}).

\bibitem[{\citenamefont{Gor}(2014)}]{Gor2014}
\bibinfo{author}{\bibfnamefont{G.~Y.} \bibnamefont{Gor}},
  \bibinfo{journal}{Langmuir} \textbf{\bibinfo{volume}{30}},
  \bibinfo{pages}{13564} (\bibinfo{year}{2014}).

\bibitem[{\citenamefont{Norman and Filinov}(1969)}]{Norman1969}
\bibinfo{author}{\bibfnamefont{G.~E.} \bibnamefont{Norman}} \bibnamefont{and}
  \bibinfo{author}{\bibfnamefont{V.~S.} \bibnamefont{Filinov}},
  \bibinfo{journal}{High Temp.} \textbf{\bibinfo{volume}{7}},
  \bibinfo{pages}{216} (\bibinfo{year}{1969}).

\bibitem[{\citenamefont{Frenkel and Smit}(2001)}]{Frenkel-Smit}
\bibinfo{author}{\bibfnamefont{D.}~\bibnamefont{Frenkel}} \bibnamefont{and}
  \bibinfo{author}{\bibfnamefont{B.}~\bibnamefont{Smit}},
  \emph{\bibinfo{title}{Understanding molecular simulation: from algorithms to
  applications}}, vol.~\bibinfo{volume}{1} (\bibinfo{publisher}{Academic
  press}, \bibinfo{year}{2001}).

\bibitem[{\citenamefont{Siderius and Shen}(2013)}]{Siderius2013}
\bibinfo{author}{\bibfnamefont{D.~W.} \bibnamefont{Siderius}} \bibnamefont{and}
  \bibinfo{author}{\bibfnamefont{V.~K.} \bibnamefont{Shen}},
  \bibinfo{journal}{J. Phys. Chem. C} \textbf{\bibinfo{volume}{117}},
  \bibinfo{pages}{5861} (\bibinfo{year}{2013}).

\bibitem[{\citenamefont{Thommes and Cychosz}(2014)}]{Thommes2014}
\bibinfo{author}{\bibfnamefont{M.}~\bibnamefont{Thommes}} \bibnamefont{and}
  \bibinfo{author}{\bibfnamefont{K.~A.} \bibnamefont{Cychosz}},
  \bibinfo{journal}{Adsorption} \textbf{\bibinfo{volume}{20}},
  \bibinfo{pages}{233} (\bibinfo{year}{2014}).

\bibitem[{\citenamefont{Brodskaya
  et~al.}(2010{\natexlab{a}})\citenamefont{Brodskaya, Rusanov, and
  Kuni}}]{Brodskaya2010i}
\bibinfo{author}{\bibfnamefont{E.~N.} \bibnamefont{Brodskaya}},
  \bibinfo{author}{\bibfnamefont{A.~I.} \bibnamefont{Rusanov}},
  \bibnamefont{and} \bibinfo{author}{\bibfnamefont{F.~M.} \bibnamefont{Kuni}},
  \bibinfo{journal}{Colloid J.} \textbf{\bibinfo{volume}{72}},
  \bibinfo{pages}{602} (\bibinfo{year}{2010}{\natexlab{a}}).

\bibitem[{\citenamefont{Brodskaya
  et~al.}(2010{\natexlab{b}})\citenamefont{Brodskaya, Rusanov, and
  Kuni}}]{Brodskaya2010ii}
\bibinfo{author}{\bibfnamefont{E.~N.}~\bibnamefont{Brodskaya}},
  \bibinfo{author}{\bibfnamefont{A.~I.}~\bibnamefont{Rusanov}}, \bibnamefont{and}
  \bibinfo{author}{\bibfnamefont{F.~M.}~\bibnamefont{Kuni}},
  \bibinfo{journal}{Colloid J.} \textbf{\bibinfo{volume}{72}},
  \bibinfo{pages}{612} (\bibinfo{year}{2010}{\natexlab{b}}).

\bibitem[{\citenamefont{Long et~al.}(2011)\citenamefont{Long, Palmer, Coasne,
  {\'S}liwinska-Bartkowiak, and Gubbins}}]{Long2011}
\bibinfo{author}{\bibfnamefont{Y.}~\bibnamefont{Long}},
  \bibinfo{author}{\bibfnamefont{J.~C.} \bibnamefont{Palmer}},
  \bibinfo{author}{\bibfnamefont{B.}~\bibnamefont{Coasne}},
  \bibinfo{author}{\bibfnamefont{M.}~\bibnamefont{{\'S}liwinska-Bartkowiak}},
  \bibnamefont{and} \bibinfo{author}{\bibfnamefont{K.~E.}
  \bibnamefont{Gubbins}}, \bibinfo{journal}{Phys. Chem. Chem. Phys.}
  \textbf{\bibinfo{volume}{13}}, \bibinfo{pages}{17163} (\bibinfo{year}{2011}).

\bibitem[{\citenamefont{Long et~al.}(2013{\natexlab{a}})\citenamefont{Long,
  Palmer, Coasne, \`Sliwinska-Bartkowiak, Jackson, M\"uller, and
  Gubbins}}]{Long2013}
\bibinfo{author}{\bibfnamefont{Y.}~\bibnamefont{Long}},
  \bibinfo{author}{\bibfnamefont{J.~C.} \bibnamefont{Palmer}},
  \bibinfo{author}{\bibfnamefont{B.}~\bibnamefont{Coasne}},
  \bibinfo{author}{\bibfnamefont{M.}~\bibnamefont{\`Sliwinska-Bartkowiak}},
  \bibinfo{author}{\bibfnamefont{G.}~\bibnamefont{Jackson}},
  \bibinfo{author}{\bibfnamefont{E.~A.} \bibnamefont{M\"uller}},
  \bibnamefont{and} \bibinfo{author}{\bibfnamefont{K.~E.}
  \bibnamefont{Gubbins}}, \bibinfo{journal}{J. Chem. Phys.}
  \textbf{\bibinfo{volume}{139}}, \bibinfo{pages}{144701}
  (\bibinfo{year}{2013}{\natexlab{a}}).

\bibitem[{\citenamefont{Long et~al.}(2013{\natexlab{b}})\citenamefont{Long,
  {\'S}liwi{\'n}ska-Bartkowiak, Drozdowski, Kempi{\'n}ski, Phillips, Palmer,
  and Gubbins}}]{Long2013Colloids}
\bibinfo{author}{\bibfnamefont{Y.}~\bibnamefont{Long}},
  \bibinfo{author}{\bibfnamefont{M.}~\bibnamefont{{\'S}liwi{\'n}ska-Bartkowiak}},
  \bibinfo{author}{\bibfnamefont{H.}~\bibnamefont{Drozdowski}},
  \bibinfo{author}{\bibfnamefont{M.}~\bibnamefont{Kempi{\'n}ski}},
  \bibinfo{author}{\bibfnamefont{K.~A.} \bibnamefont{Phillips}},
  \bibinfo{author}{\bibfnamefont{J.~C.} \bibnamefont{Palmer}},
  \bibnamefont{and} \bibinfo{author}{\bibfnamefont{K.~E.}
  \bibnamefont{Gubbins}}, \bibinfo{journal}{Colloids Surf., A}
  \textbf{\bibinfo{volume}{437}}, \bibinfo{pages}{33}
  (\bibinfo{year}{2013}{\natexlab{b}}).

\bibitem[{\citenamefont{Landau and Lifshitz}(1980)}]{Landau5}
\bibinfo{author}{\bibfnamefont{L.~D.} \bibnamefont{Landau}} \bibnamefont{and}
  \bibinfo{author}{\bibfnamefont{E.~M.} \bibnamefont{Lifshitz}},
  \emph{\bibinfo{title}{Statistical Physics, vol. 5}},
  vol.~\bibinfo{volume}{30} (\bibinfo{publisher}{Pergamon},
  \bibinfo{year}{1980}).

\bibitem[{\citenamefont{Allen and Tildesley}(1989)}]{Allen1989}
\bibinfo{author}{\bibfnamefont{M.}~\bibnamefont{Allen}} \bibnamefont{and}
  \bibinfo{author}{\bibfnamefont{D.}~\bibnamefont{Tildesley}},
  \emph{\bibinfo{title}{Computer simulation of liquids. 1987}}, vol.
  \bibinfo{volume}{385} (\bibinfo{publisher}{New York: Oxford},
  \bibinfo{year}{1989}).

\bibitem[{\citenamefont{Metropolis et~al.}(1953)\citenamefont{Metropolis,
  Rosenbluth, Rosenbluth, Teller, and Teller}}]{Metropolis1953}
\bibinfo{author}{\bibfnamefont{N.}~\bibnamefont{Metropolis}},
  \bibinfo{author}{\bibfnamefont{A.~W.} \bibnamefont{Rosenbluth}},
  \bibinfo{author}{\bibfnamefont{M.~N.} \bibnamefont{Rosenbluth}},
  \bibinfo{author}{\bibfnamefont{A.~H.} \bibnamefont{Teller}},
  \bibnamefont{and} \bibinfo{author}{\bibfnamefont{E.}~\bibnamefont{Teller}},
  \bibinfo{journal}{J. Chem. Phys.} \textbf{\bibinfo{volume}{21}},
  \bibinfo{pages}{1087} (\bibinfo{year}{1953}).

\bibitem[{\citenamefont{Gubbins et~al.}(2011)\citenamefont{Gubbins, Liu, Moore,
  and Palmer}}]{Gubbins2011}
\bibinfo{author}{\bibfnamefont{K.~E.} \bibnamefont{Gubbins}},
  \bibinfo{author}{\bibfnamefont{Y.-C.} \bibnamefont{Liu}},
  \bibinfo{author}{\bibfnamefont{J.~D.} \bibnamefont{Moore}}, \bibnamefont{and}
  \bibinfo{author}{\bibfnamefont{J.~C.} \bibnamefont{Palmer}},
  \bibinfo{journal}{Phys. Chem. Chem. Phys.} \textbf{\bibinfo{volume}{13}},
  \bibinfo{pages}{58} (\bibinfo{year}{2011}).

\bibitem[{\citenamefont{Baksh and Yang}(1991)}]{Baksh1991}
\bibinfo{author}{\bibfnamefont{M.}~\bibnamefont{Baksh}} \bibnamefont{and}
  \bibinfo{author}{\bibfnamefont{R.}~\bibnamefont{Yang}},
  \bibinfo{journal}{AIChE J.} \textbf{\bibinfo{volume}{37}},
  \bibinfo{pages}{923} (\bibinfo{year}{1991}).

\bibitem[{\citenamefont{Ravikovitch and Neimark}(2002)}]{Ravikovitch2002}
\bibinfo{author}{\bibfnamefont{P.~I.} \bibnamefont{Ravikovitch}}
  \bibnamefont{and} \bibinfo{author}{\bibfnamefont{A.~V.}
  \bibnamefont{Neimark}}, \bibinfo{journal}{Langmuir}
  \textbf{\bibinfo{volume}{18}}, \bibinfo{pages}{1550} (\bibinfo{year}{2002}).

\bibitem[{\citenamefont{Vishnyakov and Neimark}(2001)}]{Vishnyakov2001}
\bibinfo{author}{\bibfnamefont{A.}~\bibnamefont{Vishnyakov}} \bibnamefont{and}
  \bibinfo{author}{\bibfnamefont{A.~V.} \bibnamefont{Neimark}},
  \bibinfo{journal}{J. Phys. Chem. B} \textbf{\bibinfo{volume}{105}},
  \bibinfo{pages}{7009} (\bibinfo{year}{2001}).

\bibitem[{\citenamefont{Ravikovitch et~al.}(1997)\citenamefont{Ravikovitch,
  Wei, Chueh, Haller, and Neimark}}]{Ravikovitch1997}
\bibinfo{author}{\bibfnamefont{P.~I.} \bibnamefont{Ravikovitch}},
  \bibinfo{author}{\bibfnamefont{D.}~\bibnamefont{Wei}},
  \bibinfo{author}{\bibfnamefont{W.~T.} \bibnamefont{Chueh}},
  \bibinfo{author}{\bibfnamefont{G.~L.} \bibnamefont{Haller}},
  \bibnamefont{and} \bibinfo{author}{\bibfnamefont{A.~V.}
  \bibnamefont{Neimark}}, \bibinfo{journal}{J. Phys. Chem. B}
  \textbf{\bibinfo{volume}{101}}, \bibinfo{pages}{3671} (\bibinfo{year}{1997}).

\bibitem[{\citenamefont{Rasmussen et~al.}(2010)\citenamefont{Rasmussen,
  Vishnyakov, Thommes, Smarsly, Kleitz, and Neimark}}]{Rasmussen2010}
\bibinfo{author}{\bibfnamefont{C.~J.} \bibnamefont{Rasmussen}},
  \bibinfo{author}{\bibfnamefont{A.}~\bibnamefont{Vishnyakov}},
  \bibinfo{author}{\bibfnamefont{M.}~\bibnamefont{Thommes}},
  \bibinfo{author}{\bibfnamefont{B.~M.} \bibnamefont{Smarsly}},
  \bibinfo{author}{\bibfnamefont{F.}~\bibnamefont{Kleitz}}, \bibnamefont{and}
  \bibinfo{author}{\bibfnamefont{A.~V.} \bibnamefont{Neimark}},
  \bibinfo{journal}{Langmuir} \textbf{\bibinfo{volume}{26}},
  \bibinfo{pages}{10147} (\bibinfo{year}{2010}).

\bibitem[{\citenamefont{Gor et~al.}(2012)\citenamefont{Gor, Rasmussen, and
  Neimark}}]{Gor2012}
\bibinfo{author}{\bibfnamefont{G.~Y.} \bibnamefont{Gor}},
  \bibinfo{author}{\bibfnamefont{C.~J.} \bibnamefont{Rasmussen}},
  \bibnamefont{and} \bibinfo{author}{\bibfnamefont{A.~V.}
  \bibnamefont{Neimark}}, \bibinfo{journal}{Langmuir}
  \textbf{\bibinfo{volume}{28}}, \bibinfo{pages}{12100} (\bibinfo{year}{2012}).

\bibitem[{\citenamefont{Johnson et~al.}(1993)\citenamefont{Johnson, Zollweg,
  and Gubbins}}]{Johnson1993}
\bibinfo{author}{\bibfnamefont{J.~K.} \bibnamefont{Johnson}},
  \bibinfo{author}{\bibfnamefont{J.~A.} \bibnamefont{Zollweg}},
  \bibnamefont{and} \bibinfo{author}{\bibfnamefont{K.~E.}
  \bibnamefont{Gubbins}}, \bibinfo{journal}{Mol. Phys.}
  \textbf{\bibinfo{volume}{78}}, \bibinfo{pages}{591} (\bibinfo{year}{1993}).

\bibitem[{Not({\natexlab{a}})}]{Note1}
\bibinfo{note}{To estimate the errors for the calculated compressibility, for
  each of the data sets (fluctuating number of molecules $N$) we calculated the autocorrelation function (AC) $c_i$ at lag $i$. From $\left\lbrace c_i \right\rbrace$  the ``decorrelation time'' is calculated as
  (Eq.~(2.16) in Ref.~\onlinecite{Sokal1997}) $\tau = 1 + \frac{2}{c_0}
  \sum\limits_{i=1}^{t_0} c_i,$ where $t_0$ is a characteristic time at which
  AC becomes close to zero. Since this formalism simply corrects for
  correlations between configuration, in MC simulations the number of attempted
  MC moves can be considered an analogue of the discrete time
  variable $t$. The effective number of uncorrelated steps is given then as
  $t_{\rm{eff}} = t_{\rm{total}}/\tau$. It allows us to estimate the
  relative error as $\sqrt{\frac{2}{t_{\rm{eff}} - 1}}.$ \cite{Lehmann1998} }. 

\bibitem[{\citenamefont{Errington}(2003{\natexlab{a}})}]{Errington_Evaluating_2003}
\bibinfo{author}{\bibfnamefont{J.~R.} \bibnamefont{Errington}},
  \bibinfo{journal}{Phys. Rev. E} \textbf{\bibinfo{volume}{67}},
  \bibinfo{pages}{012102} (\bibinfo{year}{2003}{\natexlab{a}}).

\bibitem[{\citenamefont{Errington}(2003{\natexlab{b}})}]{Errington_Direct_2003}
\bibinfo{author}{\bibfnamefont{J.~R.} \bibnamefont{Errington}},
  \bibinfo{journal}{J. Chem. Phys.} \textbf{\bibinfo{volume}{118}},
  \bibinfo{pages}{9915} (\bibinfo{year}{2003}{\natexlab{b}}).

\bibitem[{\citenamefont{Shen and Errington}(2004)}]{Shen_Metastability_2004}
\bibinfo{author}{\bibfnamefont{V.~K.} \bibnamefont{Shen}} \bibnamefont{and}
  \bibinfo{author}{\bibfnamefont{J.~R.} \bibnamefont{Errington}},
  \bibinfo{journal}{J. Phys. Chem. B} \textbf{\bibinfo{volume}{108}},
  \bibinfo{pages}{19595} (\bibinfo{year}{2004}).

\bibitem[{\citenamefont{Shen and Errington}(2005)}]{Shen_Determination_2005}
\bibinfo{author}{\bibfnamefont{V.~K.} \bibnamefont{Shen}} \bibnamefont{and}
  \bibinfo{author}{\bibfnamefont{J.~R.} \bibnamefont{Errington}},
  \bibinfo{journal}{J. Chem. Phys.} \textbf{\bibinfo{volume}{122}},
  \bibinfo{pages}{064508} (\bibinfo{year}{2005}).

\bibitem[{\citenamefont{Errington and Shen}(2005)}]{Errington_Direct_2005}
\bibinfo{author}{\bibfnamefont{J.~R.} \bibnamefont{Errington}}
  \bibnamefont{and} \bibinfo{author}{\bibfnamefont{V.~K.} \bibnamefont{Shen}},
  \bibinfo{journal}{J. Chem. Phys.} \textbf{\bibinfo{volume}{123}},
  \bibinfo{pages}{164103} (\bibinfo{year}{2005}).

\bibitem[{\citenamefont{Shen and Errington}(2006)}]{Shen_Determination_2006}
\bibinfo{author}{\bibfnamefont{V.~K.} \bibnamefont{Shen}} \bibnamefont{and}
  \bibinfo{author}{\bibfnamefont{J.~R.} \bibnamefont{Errington}},
  \bibinfo{journal}{J. Chem. Phys.} \textbf{\bibinfo{volume}{124}},
  \bibinfo{pages}{024721} (\bibinfo{year}{2006}).

\bibitem[{\citenamefont{Shen and Siderius}(2014)}]{Shen_Elucidating_2014}
\bibinfo{author}{\bibfnamefont{V.~K.} \bibnamefont{Shen}} \bibnamefont{and}
  \bibinfo{author}{\bibfnamefont{D.~W.} \bibnamefont{Siderius}},
  \bibinfo{journal}{J. Chem. Phys.} \textbf{\bibinfo{volume}{104}},
  \bibinfo{pages}{244106} (\bibinfo{year}{2014}).

\bibitem[{\citenamefont{Shell et~al.}(2003)\citenamefont{Shell, Debenedetti,
  and Panagiotopoulos}}]{Shell_improved_2003}
\bibinfo{author}{\bibfnamefont{M.~S.} \bibnamefont{Shell}},
  \bibinfo{author}{\bibfnamefont{P.~G.} \bibnamefont{Debenedetti}},
  \bibnamefont{and} \bibinfo{author}{\bibfnamefont{A.~Z.}
  \bibnamefont{Panagiotopoulos}}, \bibinfo{journal}{J. Chem. Phys.}
  \textbf{\bibinfo{volume}{119}}, \bibinfo{pages}{9406} (\bibinfo{year}{2003}).

\bibitem[{\citenamefont{Rane et~al.}(2013)\citenamefont{Rane, Murali, and
  Errington}}]{Rane_Monte_2013}
\bibinfo{author}{\bibfnamefont{K.~S.} \bibnamefont{Rane}},
  \bibinfo{author}{\bibfnamefont{S.}~\bibnamefont{Murali}}, \bibnamefont{and}
  \bibinfo{author}{\bibfnamefont{J.~R.} \bibnamefont{Errington}},
  \bibinfo{journal}{J. Chem. Theory Comput.} \textbf{\bibinfo{volume}{9}},
  \bibinfo{pages}{2552} (\bibinfo{year}{2013}).

\bibitem[{\citenamefont{Lyubartsev et~al.}(1992)\citenamefont{Lyubartsev,
  Martsinovski, Shevkunov, and Vorontsov-Velyaminov}}]{Lyubartsev_New_1991}
\bibinfo{author}{\bibfnamefont{A.~P.} \bibnamefont{Lyubartsev}},
  \bibinfo{author}{\bibfnamefont{A.~A.} \bibnamefont{Martsinovski}},
  \bibinfo{author}{\bibfnamefont{S.~V.} \bibnamefont{Shevkunov}},
  \bibnamefont{and} \bibinfo{author}{\bibfnamefont{P.~N.}
  \bibnamefont{Vorontsov-Velyaminov}}, \bibinfo{journal}{J. Chem. Phys.}
  \textbf{\bibinfo{volume}{96}}, \bibinfo{pages}{1776} (\bibinfo{year}{1992}).

\bibitem[{\citenamefont{Escobedo and de~Pablo
  J.~J.}(1996)}]{Escobedo_Expanded_1996}
\bibinfo{author}{\bibfnamefont{F.~A.} \bibnamefont{Escobedo}} \bibnamefont{and}
  \bibinfo{author}{\bibnamefont{de~Pablo J.~J.}}, \bibinfo{journal}{J. Chem.
  Phys.} \textbf{\bibinfo{volume}{105}}, \bibinfo{pages}{4931}
  (\bibinfo{year}{1996}).

\bibitem[{\citenamefont{Errington}(2004)}]{Errington_Prewetting_2004}
\bibinfo{author}{\bibfnamefont{J.~R.} \bibnamefont{Errington}},
  \bibinfo{journal}{Langmuir} \textbf{\bibinfo{volume}{20}},
  \bibinfo{pages}{3798} (\bibinfo{year}{2004}).

\bibitem[{\citenamefont{Kuni}(1981)}]{Kuni1981}
\bibinfo{author}{\bibfnamefont{F.~M.} \bibnamefont{Kuni}},
  \emph{\bibinfo{title}{Statistical physics and thermodynamics}}
  (\bibinfo{publisher}{Moscow, Nauka}, \bibinfo{year}{1981}).

\bibitem[{\citenamefont{Tegeler et~al.}(1999)\citenamefont{Tegeler, Span, and
  Wagner}}]{Tegeler1999}
\bibinfo{author}{\bibfnamefont{C.}~\bibnamefont{Tegeler}},
  \bibinfo{author}{\bibfnamefont{R.}~\bibnamefont{Span}}, \bibnamefont{and}
  \bibinfo{author}{\bibfnamefont{W.}~\bibnamefont{Wagner}},
  \bibinfo{journal}{J. Phys. Chem. Ref. Data} \textbf{\bibinfo{volume}{28}},
  \bibinfo{pages}{779} (\bibinfo{year}{1999}).

\bibitem[{Not({\natexlab{b}})}]{Note2}
\bibinfo{note}{When discussing the dependence on $p/p_0$, the modulus $K_T$ is
  more convenient than $\beta _T$, since it is expected to be a linear function
  of $\protect \qopname \relax o{log}(p/p_0)$.}

\bibitem[{\citenamefont{Coasne et~al.}(2009)\citenamefont{Coasne, Czwartos,
  Sliwinska-Bartkowiak, and Gubbins}}]{Coasne2009}
\bibinfo{author}{\bibfnamefont{B.}~\bibnamefont{Coasne}},
  \bibinfo{author}{\bibfnamefont{J.}~\bibnamefont{Czwartos}},
  \bibinfo{author}{\bibfnamefont{M.}~\bibnamefont{Sliwinska-Bartkowiak}},
  \bibnamefont{and} \bibinfo{author}{\bibfnamefont{K.~E.}
  \bibnamefont{Gubbins}}, \bibinfo{journal}{J. Phys. Chem. B}
  \textbf{\bibinfo{volume}{113}}, \bibinfo{pages}{13874}
  (\bibinfo{year}{2009}).

\bibitem[{\citenamefont{Erko et~al.}(2012)\citenamefont{Erko, Wallacher, Hoell,
  Hauss, Zizak, and Paris}}]{Erko2012}
\bibinfo{author}{\bibfnamefont{M.}~\bibnamefont{Erko}},
  \bibinfo{author}{\bibfnamefont{D.}~\bibnamefont{Wallacher}},
  \bibinfo{author}{\bibfnamefont{A.}~\bibnamefont{Hoell}},
  \bibinfo{author}{\bibfnamefont{T.}~\bibnamefont{Hauss}},
  \bibinfo{author}{\bibfnamefont{I.}~\bibnamefont{Zizak}}, \bibnamefont{and}
  \bibinfo{author}{\bibfnamefont{O.}~\bibnamefont{Paris}},
  \bibinfo{journal}{Phys. Chem. Chem. Phys.} \textbf{\bibinfo{volume}{14}},
  \bibinfo{pages}{3852} (\bibinfo{year}{2012}).

\bibitem[{\citenamefont{Gor and Neimark}(2010)}]{Gor2010}
\bibinfo{author}{\bibfnamefont{G.~Y.} \bibnamefont{Gor}} \bibnamefont{and}
  \bibinfo{author}{\bibfnamefont{A.~V.} \bibnamefont{Neimark}},
  \bibinfo{journal}{Langmuir} \textbf{\bibinfo{volume}{26}},
  \bibinfo{pages}{13021} (\bibinfo{year}{2010}).

\bibitem[{\citenamefont{Warner and Beamish}(1988)}]{Warner1988}
\bibinfo{author}{\bibfnamefont{K.}~\bibnamefont{Warner}} \bibnamefont{and}
  \bibinfo{author}{\bibfnamefont{J.}~\bibnamefont{Beamish}},
  \bibinfo{journal}{J. Appl. Phys.} \textbf{\bibinfo{volume}{63}},
  \bibinfo{pages}{4372} (\bibinfo{year}{1988}).

\bibitem[{\citenamefont{Barrett et~al.}(1951)\citenamefont{Barrett, Joyner, and
  Halenda}}]{Barrett1951}
\bibinfo{author}{\bibfnamefont{E.~P.} \bibnamefont{Barrett}},
  \bibinfo{author}{\bibfnamefont{L.~G.} \bibnamefont{Joyner}},
  \bibnamefont{and} \bibinfo{author}{\bibfnamefont{P.~P.}
  \bibnamefont{Halenda}}, \bibinfo{journal}{J. Am. Chem. Soc.}
  \textbf{\bibinfo{volume}{73}}, \bibinfo{pages}{373} (\bibinfo{year}{1951}).

\bibitem[{\citenamefont{Landers et~al.}(2013)\citenamefont{Landers, Gor, and
  Neimark}}]{Landers2013}
\bibinfo{author}{\bibfnamefont{J.}~\bibnamefont{Landers}},
  \bibinfo{author}{\bibfnamefont{G.~Y.} \bibnamefont{Gor}}, \bibnamefont{and}
  \bibinfo{author}{\bibfnamefont{A.~V.} \bibnamefont{Neimark}},
  \bibinfo{journal}{Colloids Surf., A} \textbf{\bibinfo{volume}{437}},
  \bibinfo{pages}{3} (\bibinfo{year}{2013}).

\bibitem[{\citenamefont{Dukhin et~al.}(2013)\citenamefont{Dukhin, Swasey, and
  Thommes}}]{Dukhin2013}
\bibinfo{author}{\bibfnamefont{A.}~\bibnamefont{Dukhin}},
  \bibinfo{author}{\bibfnamefont{S.}~\bibnamefont{Swasey}}, \bibnamefont{and}
  \bibinfo{author}{\bibfnamefont{M.}~\bibnamefont{Thommes}},
  \bibinfo{journal}{Colloids Surf., A} \textbf{\bibinfo{volume}{437}},
  \bibinfo{pages}{127} (\bibinfo{year}{2013}).

\bibitem[{\citenamefont{Mosher et~al.}(2013)\citenamefont{Mosher, He, Liu,
  Rupp, and Wilcox}}]{Mosher2013}
\bibinfo{author}{\bibfnamefont{K.}~\bibnamefont{Mosher}},
  \bibinfo{author}{\bibfnamefont{J.}~\bibnamefont{He}},
  \bibinfo{author}{\bibfnamefont{Y.}~\bibnamefont{Liu}},
  \bibinfo{author}{\bibfnamefont{E.}~\bibnamefont{Rupp}}, \bibnamefont{and}
  \bibinfo{author}{\bibfnamefont{J.}~\bibnamefont{Wilcox}},
  \bibinfo{journal}{Int. J. Coal Geol.} \textbf{\bibinfo{volume}{109}},
  \bibinfo{pages}{36} (\bibinfo{year}{2013}).

\bibitem[{\citenamefont{Hu et~al.}(2015)\citenamefont{Hu, Devegowda, Striolo,
  Phan, Ho, Civan, and Sigal}}]{Hu2015}
\bibinfo{author}{\bibfnamefont{Y.}~\bibnamefont{Hu}},
  \bibinfo{author}{\bibfnamefont{D.}~\bibnamefont{Devegowda}},
  \bibinfo{author}{\bibfnamefont{A.}~\bibnamefont{Striolo}},
  \bibinfo{author}{\bibfnamefont{A.}~\bibnamefont{Phan}},
  \bibinfo{author}{\bibfnamefont{T.~A.} \bibnamefont{Ho}},
  \bibinfo{author}{\bibfnamefont{F.}~\bibnamefont{Civan}}, \bibnamefont{and}
  \bibinfo{author}{\bibfnamefont{R.}~\bibnamefont{Sigal}}, \bibinfo{journal}{J.
  Unconv. Oil Gas Resour.} \textbf{\bibinfo{volume}{9}}, \bibinfo{pages}{31}
  (\bibinfo{year}{2015}).

\bibitem[{\citenamefont{Falk et~al.}(2015)\citenamefont{Falk, Coasne, Pellenq,
  Ulm, and Bocquet}}]{Falk2015}
\bibinfo{author}{\bibfnamefont{K.}~\bibnamefont{Falk}},
  \bibinfo{author}{\bibfnamefont{B.}~\bibnamefont{Coasne}},
  \bibinfo{author}{\bibfnamefont{R.}~\bibnamefont{Pellenq}},
  \bibinfo{author}{\bibfnamefont{F.-J.} \bibnamefont{Ulm}}, \bibnamefont{and}
  \bibinfo{author}{\bibfnamefont{L.}~\bibnamefont{Bocquet}},
  \bibinfo{journal}{Nat. Commun.} \textbf{\bibinfo{volume}{6}},
  \bibinfo{pages}{6949} (\bibinfo{year}{2015}).

\bibitem[{\citenamefont{Sokal}(1997)}]{Sokal1997}
\bibinfo{author}{\bibfnamefont{A.~D.} \bibnamefont{Sokal}},
  \emph{\bibinfo{title}{Functional Integration NATO ASI Series Volume 361}}
  (\bibinfo{publisher}{Springer}, \bibinfo{year}{1997}), chap.
  \bibinfo{chapter}{Monte Carlo methods in statistical mechanics: foundations
  and new algorithms}, pp. \bibinfo{pages}{131--192}.

\bibitem[{\citenamefont{Lehmann and Casella}(1998)}]{Lehmann1998}
\bibinfo{author}{\bibfnamefont{E.~L.} \bibnamefont{Lehmann}} \bibnamefont{and}
  \bibinfo{author}{\bibfnamefont{G.}~\bibnamefont{Casella}},
  \emph{\bibinfo{title}{Theory of point estimation}}, vol.~\bibinfo{volume}{31}
  (\bibinfo{publisher}{Springer Science \& Business Media},
  \bibinfo{year}{1998}).

\end{thebibliography}

\end{document}